# Building Barricades: Red-Teaming Agentic Biosynthesis Planning

Nicholas Roehner, Bryan Bartley, Jacob Beal, Caitlin Jagla, Troy Timmerman, and Kemper Talley

RTX BBN Technologies

## 1 INTRODUCTION

Biosecurity has been aided by the inherent difficulty in accessing most dangerous pathogens and toxins. As increasing capabilities in biotechnology and synthetic biology have made bioengineering more accessible, these barriers to access have been falling. The advent of LLMs is beginning to lower some of the technical barriers that have historically limited biosynthetic design to experts. In the closely related area of chemical synthesis, it has recently been demonstrated that non-expert users can leverage agentic systems to synthesize chemical warfare agents such as phosgene using only text-based LLM queries and commercial-off-the-shelf (COTS) laboratory equipment, such as Opentrons, to carry out the necessary chemical reactions (Boiko et al. 2023, Bran et al. 2024). Our present work in developing an agentic system for biosynthesis planning suggests that, as LLM-driven tools mature, the capability to design biosynthesis pathways could become more widely accessible (i.e., chemical synthesis via biological enzymes). Because biological systems can be engineered in many ways, the potential threat landscape for the creation of existing and previously unknown bioweapons and biosynthetic toxins is expansive (Urbina et al. 2022, Millett et al. 2023). The biosecurity community should consider accelerating the development of mitigation strategies to prevent the realization of these potential threats. In this work, we discuss our research into developing an agentic system for biosynthesis planning and its associated technical barriers, the implications of such systems for biosecurity, and our recommendations for mitigating the risks that they pose.

Currently, biosynthesis planning is accomplished with software requiring significant domain expertise in biology, bioinformatics, and genetic engineering (Appleton et al. 2017, Gricourt et al. 2024). Non-experts are hampered by knowledge barriers associated with tasks such as identifying the enzymes needed to synthesize a target small molecule (retrobiosynthesis) and designing gene clusters to express these enzymes. However, as the democratization of software development demonstrates, it is increasingly possible for non-experts to generate sophisticated work products without the involvement of domain experts (Joel et al.

2025). We believe a parallel phenomenon is emerging in biotechnology, one that will enable non-expert users to perform biosynthesis planning using simple natural language prompts and LLM agents that have been interfaced with biodesign tools. If knowledge barriers continue to erode, there is a plausible risk that individuals without formal training could attempt to design and implement biosynthesis pathways for small molecule toxins.

## 2 CASE STUDY: RED-TEAMING LLM-DRIVEN BIOSYNTHESIS

To assess the degree to which LLM agents could lower knowledge barriers to biosynthesis for non-experts, BBN developed an agentic system, BioBulbul, that ties together LLMs with tools for biosynthesis planning and gene cluster design (Figure 1). BioBulbul builds upon Bran et al.'s work on the ChemCrow agentic system for chemical retrosynthesis (Bran et al. 2023). Like ChemCrow, BioBulbul leverages chain-of-thought prompting to dynamically compose and execute domain-specific software tools, a form of retrieval-augmented generation (RAG) to enhance the LLM's response with highly specific biochemical and genetic information.

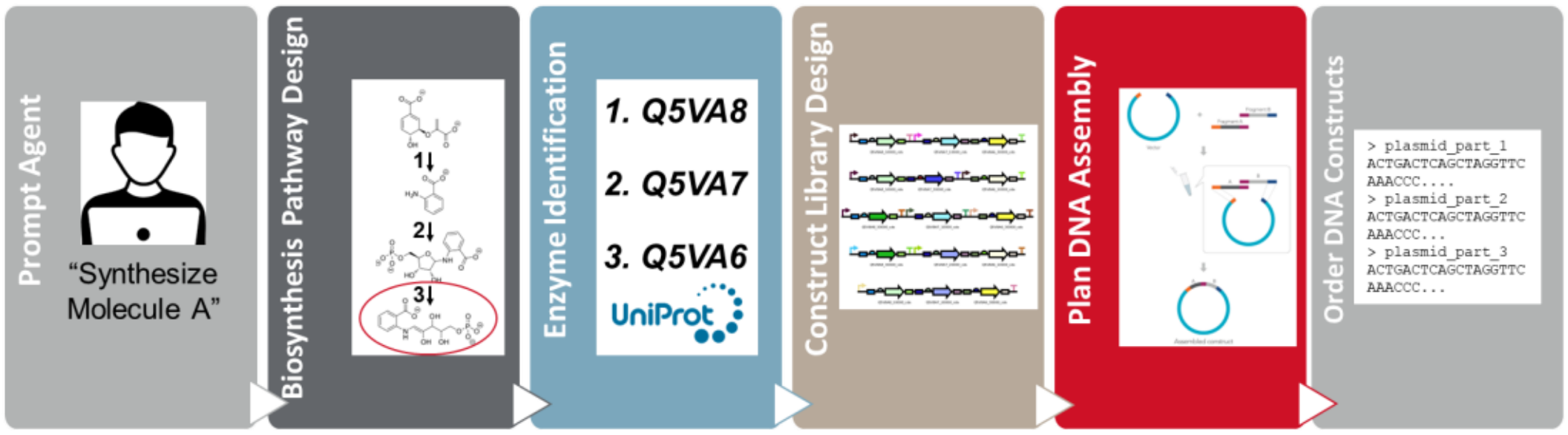

**Figure 1. The BioBulbul agent-driven workflow enables a non-expert user to perform biosynthesis planning and gene cluster design based on simple natural language prompts.**

In this context, we defined a non-expert as someone with an introductory understanding of biology and laboratory techniques: for example, a high school or undergraduate member of an International Genetically Engineered Machine (iGEM) competition team (Kelwick et al. 2015). Such an individual would at a minimum be capable of using software with a graphical user interface (GUI) and following basic protocols for genetic engineering, as taught in many high school and college laboratory classes. Our projections suggest that, within the next decade, advances could allow individuals with minimal laboratory experience to engage with biodesign tools.

For red-teaming purposes, we used publicly available, open-source tools in BioBulbul as much as possible. BioBulbul incorporates different types of tools for planning biosynthesis pathways (Figure 2). By working backwards, these planning tools identify series of enzymatic reactions that can produce target compounds from simpler precursors that are commercially available or naturally available as part of a host organism's metabolism. This computational process is called retrobiosynthesis. However, each planning tool employs different retrobiosynthesis algorithms, with a particularly important distinction between template-based and template-free planning. Template-based planners (Delépine et al. 2018) use interpretable reaction rules designed by experts or extracted from databases. Template-free planners (Tu et al. 2025) use black-box models trained on reaction databases, introducing another layer of AI-enabled design.

For designing gene clusters to implement planned biosynthesis pathways, we integrated tools from the Galaxy SynBioCAD pipeline (Hérisson et al. 2022) into BioBulbul. These include biodesign software for:

1. Selecting specific enzymes to implement planned biosynthesis reactions.
2. Designing genes to express these enzymes and designing libraries of gene clusters that vary enzyme expression using different genetic parts such as promoters and ribosome binding sites.
3. Planning the assembly of these biosynthesis gene clusters from smaller DNA fragments into constructs ready for expression in a host organism.

With BioBulbul, a simple natural language prompt in the form of "Synthesize [compound name]" is sufficient for the LLM agent to select and invoke tools for biosynthesis planning followed by genetic design (Figure 1). The final result is a single spreadsheet with sequences for synthesizable DNA fragments that are compatible with a standard Gibson assembly protocol (which LLMs are presently capable of generating as well).

As part of a red-teaming exercise, we confirmed that BioBulbul is capable of executing this workflow for several known small molecule toxins based on user prompts, generating computational results that qualitatively agree with our understanding of their known biosynthesis pathways and of what potentially viable gene clusters for implementing these pathways might look like. This demonstrates the potential of agentic systems to lower existing knowledge barriers to these and potentially other small molecule toxins. For safety and responsible disclosure, output designs are not included in this manuscript. Although their biosynthesis pathways are publicly known, the gene cluster designs generated by BioBulbul are not, and efforts are ongoing to deploy additional safeguards against their biomanufacturing. All work described here was performed in a controlled, secure environment and is not intended as a protocol for independent replication.

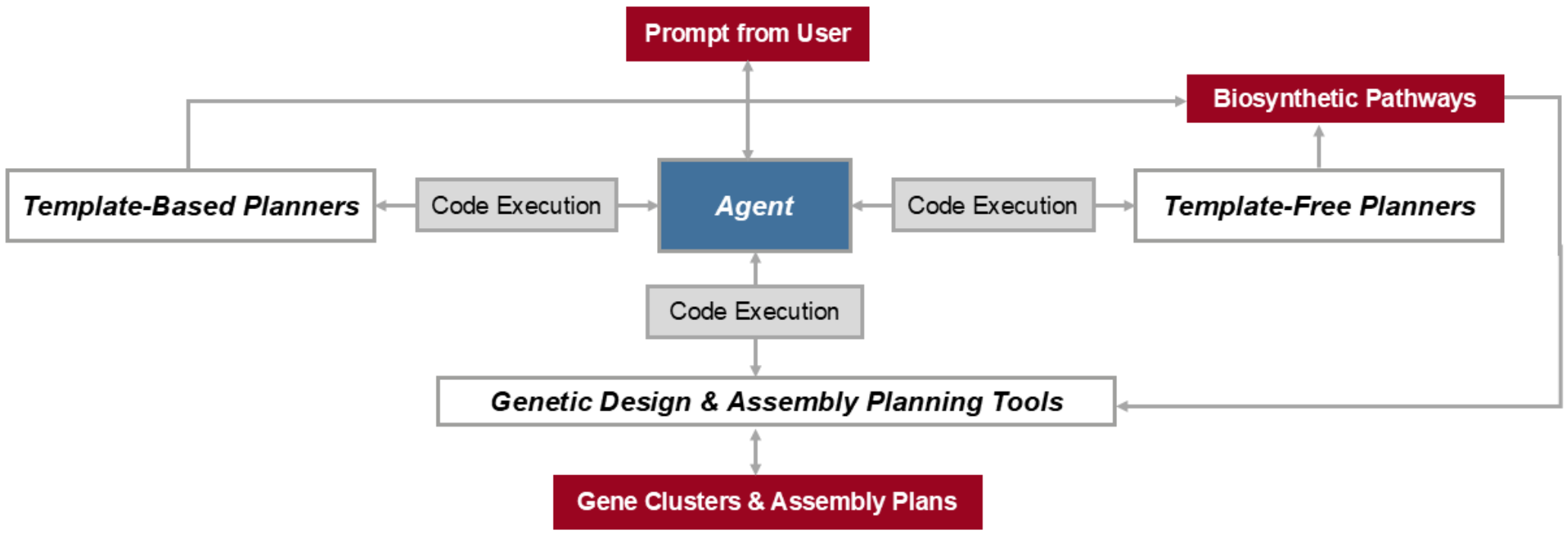

**Figure 2. The Biobulbul agentic system includes both template-based and template-free planners capable of discovering non-obvious pathways for biosynthesis that constitute potential blind spots in DNA synthesis screening.**

By including both template-based and template-free planning tools, BioBulbul may eventually uncover non-obvious pathways for synthesizing a molecule that constitute potential blind spots in screening that experts may overlook. These planner types allow the generation of two categories of biosynthesis pathways: those composed from known enzymatic reaction steps and those consisting of hypothetical reaction steps based on model predictions. For the former, template-based planners identify an exact list of known enzymatic reactions that may be introduced into a host organism, thereby extending the host's native metabolism to produce the target molecule. These pathways may recapitulate parts of actual biosynthesis pathways that exist in nature or may represent novel, alternative biochemical pathways that are less obvious. Template-free planners are capable of revealing even less obvious biosynthesis pathways. These non-obvious pathways may consist of reactions that are as of yet undiscovered and uncharacterized but are nevertheless deemed to be feasible by a model based on similarity in features to known reactions. The degree to which such novel pathways can be implemented in practice remains underexplored.

# 3 BIOSECURITY DISCUSSION AND OUTLOOK

## 3.1 Current Technical Barriers

Based on our experiences in developing BioBulbul and our assessment of the known landscape of biothreats, there are at least three barriers standing in the way of LLM-enabled biosynthesis of small molecule toxins. The first and arguably least consequential of these barriers is the relative dearth of freely available open-source biodesign tools for biosynthesis planning and gene cluster design (Gricourt et al.

2024). Additionally, many of the available tools exhibit design and/or documentation flaws that inhibit non-experts' ability to integrate them into an agentic system like BioBulbul. This particular barrier, however, will eventually be eroded by the release of new or updated open-source biodesign tools, and once it ceases to be a barrier for one group, it potentially ceases to be a barrier for all.

The second barrier to LLM-enabled toxin biosynthesis is the availability of suitable databases and training data for biodesign tools. Biosynthesis planners are particularly impacted by this issue; many rely on non-enzymatic reactions in their reference data because of limited available data on enzymatic reactions. This barrier is more difficult to overcome than the first because new research must be conducted and published before biosynthesis planners can draw upon it. Regardless, the dual-use nature of that research means it is only a matter of time before previously unplannable small molecule toxins become plannable. Further, many published enzymatic reactions are not yet included in structured databases, but increasing automation of knowledge extraction from free text (Lin et al. 2025) and images (Wilary and Cole 2023) will quickly increase the availability of structured reaction data.

The third and currently strongest barrier to LLM-enabled toxin biosynthesis is the relative complexity of many known pathways (i.e., more than six reaction steps) and the enzymes that are often involved (e.g., non-ribosomal peptide synthetases). Both factors complicate the design, construction, and implementation of a functional gene cluster capable of milligram-level toxin biosynthesis production. Our analysis, however, does not account for undiscovered toxins or pathways that may not share these challenges.

### 3.2 Practical Mitigations

Given the certain erosion of the first two barriers with time and the indeterminate robustness of the third barrier, it is prudent to consider practical mitigations for the potential threat of LLM-enabled toxin biosynthesis.

#### 3.2.1 Screening purchases for control of potential hazards

With respect to biosynthesis of small molecule toxins, control at the point of manufacture means screening DNA synthesis orders for sequences of enzymes that are more likely than not to be involved in toxin biosynthesis. Commercial DNA synthesis providers and manufacturers of benchtop nucleic acid synthesizers routinely screen sequences for other biological threats, e.g. viruses, pathogenic bacteria, and protein-based toxins, prior to actual production of any genetic material. Through our preliminary research, we have identified putative "chokepoint" enzymes within currently known biosynthesis pathways for regulated toxins as high priority candidates for screening. The Sequence Biosecurity Risk Consortium (SBRC), an industry-led group that facilitates standardization of this screening for sequences of concern, has begun to deploy safeguards based on these enzymes. We believe expanding DNA-synthesis screening to include chokepoint enzymes as an industry-wide best practice would help mitigate both known and emerging biological threats.

A second pillar of industry-wide best practices could leverage existing control strategies used by chemical suppliers and extend them to include screening purchasers of commercially available precursor compounds necessary for the biosynthesis of small molecule toxins. For direct control of single precursors with dual uses, this strategy would be limited to substrates of chokepoint enzymes. However, combinatorial screening for distinctive groupings of precursors, reagents, hardware, and/or PPE indicative of potential malicious intent could decrease risks while minimizing impact on the wider bioeconomy. Any such screening would need to be implemented in a manner consistent with FTC guidance and emerging privacy legislation, with transparent notice to customers. Techniques from financial fraud detection (Motie and Raahemi 2024) and consumer behavior analysis (Adulyasak et al. 2024), such as anomaly detection, could be adapted to flag atypical ordering patterns while respecting privacy safeguards. This would be particularly effective at flagging pattern-of-life changes in customers with ostensibly legitimate access, either in comparison to their personal norms or to the overall distribution of behavior patterns.

### 3.2.2 Sustainable threat landscape surveillance

Continuing to search for new pathways as tools and data become available is key for enduring implementation of practical mitigations, and deployment of successful controls for novel threats requires discovering them before malicious actors attempt to use them. The scale and longevity of the risk posed by the AI-driven rapid evolution of biological design tools necessitates a systematic approach to threat landscape surveillance, vulnerability assessment, and hazard identification. This approach could involve developing an agentic AI-assisted human-in-the-loop workflow to detect and assess emerging biodesign tools and datasets before vulnerabilities enabled by them can become widespread threats. Many concepts and strategies from cybersecurity could be adapted to inform the design of such a workflow (Ross et al. 2022).

Sustainable success of this strategy would be made possible in part by automating the identification of emerging risks. This includes adapting existing and emerging tools built or fine-tuned for scientific literature reviews (Dennstädt et al. 2024, Mikriukov et al. 2025) and new publication alerts, social media listening (Kitchens et al. 2024), media monitoring, website change detection, and similar tasks. Potentially hazardous biodesign tools and datasets identified by this type of continuous observation would need triaging prior to full assessment – only viable (both *distinct* and *accessible*) candidates should be funneled downstream. Biodesign tools that pass the triage gate must undergo a deeper proliferation risk analysis with a series of increasingly stringent gates designed to evaluate the likelihood of a malicious actor to successfully install, run, and produce potentially harmful outputs using that tool. These proliferation risk assessments could be at least partially managed by the agent, as BioBulbul does in the case of LLM-enabled retrobiosynthesis. However, the details of these gated assessments and definitions of thresholds used to flag vulnerabilities should be unique for each type of tool and not publicly disclosed to avoid enabling evasion.

#### 3.2.3 LLM guardrails to limit risk and enable detection

Based on the principle of least effort (Zipf 1949), bad actors may attempt to jailbreak LLMs before deploying agentic frameworks to obtain biosynthesis pathways. This provides an opportunity to develop LLM guardrails such that bad actors can be more easily identified and their harmful goals can be prevented or otherwise mitigated. For example, an LLM could be biased such that, when it is prompted to return a biosynthesis pathway for a controlled toxin, it is more likely to return pathways that are easily detected via screening due to the inclusion of controlled precursors, pathways that are more complex and thus less likely to work, or pathways that synthesize less toxic alternatives.

For these types of LLM guardrails to succeed, more research is needed to ensure that there is high sensitivity and specificity in only returning biased pathways to determined threat actors, and such that the guardrails are robust to model fine-tuning and other methods to bypass them. Collaboration with LLM developers to explore safeguard mechanisms, such as selective biasing, may be a fruitful research direction.

## 4 SUMMARY AND RECOMMENDATIONS

LLMs and agentic systems threaten to greatly lower knowledge barriers to producing small molecule toxins via biosynthesis, making it possible for non-experts to invoke relevant biodesign tools and interpret their results. While there exist technical barriers that currently prevent widespread access to LLM-driven biosynthesis, most notably the scarcity of open-source biodesign tools, limited and unstructured training data, and the relative complexity of known toxin pathways, these barriers do not appear to be insurmountable in the coming decade. Our red-team experiments indicate that, for a limited set of well-characterized toxins, LLM-driven design can generate plausible biosynthesis pathways and gene cluster sequences.

Consequently, we recommend the pursuit of practical mitigations and safeguards spanning existing routes for both biomanufacturing and biodesign. For biomanufacturing, these safeguards include expanding DNA synthesis screening pipelines to include sequences for “chokepoint” enzymes primarily involved in the biosynthesis of toxins and extending chemical-supplier controls to precursor compounds for toxin biosynthesis. For biodesign, we recommend developing automated, potentially agentic approaches that focus on continuous threat surveillance to triage emerging biodesign tools and datasets before they become widely available. We also recommend conducting further research into LLM guardrails to make malicious use of LLMs for biodesign less effective and easier to detect.

The ultimate success of these proposed mitigations will depend on continuing investment into biosecurity research and coordination between stakeholders in government, the biomanufacturing industry (including organizations such as the SBRC), and groups developing LLMs and biodesign tools. By investing in practical mitigations, engaging stakeholders, and developing standards for screening and rapid reporting, the biosecurity community can greatly reduce the likelihood that small molecule toxins are synthesized by non-expert actors using advanced AI agents.

## 5 ACKNOWLEDGEMENTS

This work was supported by the Defense Advanced Research Projects Agency (DARPA) Biological Technologies Office (BTO) under award number HR001124C0470. The views, opinions, and/or findings expressed are those of the author(s) and should not be interpreted as representing the official views or policies of DARPA or the U.S. Government. This document does not contain technology or Technical Data controlled under either the U.S. International Traffic in Arms Regulations or the U.S. Export Administration Regulations. Distribution statement A: approved for public release, distribution unlimited (DARPA DISTAR case 41923, 12/26/2025).